\documentclass{iaus}
\usepackage{graphicx}

\title[Transits and Image Deconvolution] 
{Transits against Fainter Stars:\\ The Power of Image Deconvolution}

\author[Sackett et al]   
{Penny D.\ Sackett$^1$, Micha\"el Gillon$^2$, \\ Daniel D.R.\ Bayliss$^1$, 
David T.F.\ Weldrake$^3$ \and Brandon Tingley$^4$}

\affiliation{$^1$Research School of Astronomy and Astrophysics, Australian National University,\\ Mount Stromlo, Australia\\ email: {\tt Penny.Sackett@anu.edu.au} \\[\affilskip]
$^2$Observatoire de l'Universit\'e de Gen\`eve, Switzerland\\[\affilskip]
$^3$Harvard-Smithsonian Center for Astrophysics, USA\\[\affilskip]
$^4$Instituto de Astrofisica de Canarias, Spain}

\pubyear{2009}
\volume{253}  
\pagerange{55-61}
 \date{2008}
\doi{10.1017/S1743921308026239}
\jname{Transiting Planets}
\editors{F. Pont, D. Sasselov  \& M. Holman eds.}
\begin{document}

\maketitle

\begin{abstract}
Compared to bright star searches, surveys for transiting planets against fainter 
($V=12-18$) stars have the advantage of much higher sky densities of dwarf star 
primaries, which afford easier detection of small transiting bodies.  
Furthermore, deep searches are capable of probing a wider range of stellar environments.  
On the other hand, for a given spatial resolution and transit depth, deep 
searches are more prone to confusion from blended eclipsing binaries.  
We present a powerful mitigation strategy for the blending problem 
that includes the use of image deconvolution and high resolution imaging.  
The techniques are illustrated with Lupus-TR-3 and very recent IR imaging 
with {\sc PANIC} on Magellan.  The results are likely to have implications for the 
{\sc CoRoT} and {\sc KEPLER} missions designed to detect transiting planets of 
terrestrial size.
\end{abstract}

\firstsection 
\section{Introduction}

\begin{figure}[t]
\vspace*{-0.5 cm}
\begin{center}
 \includegraphics[width=5.5in]{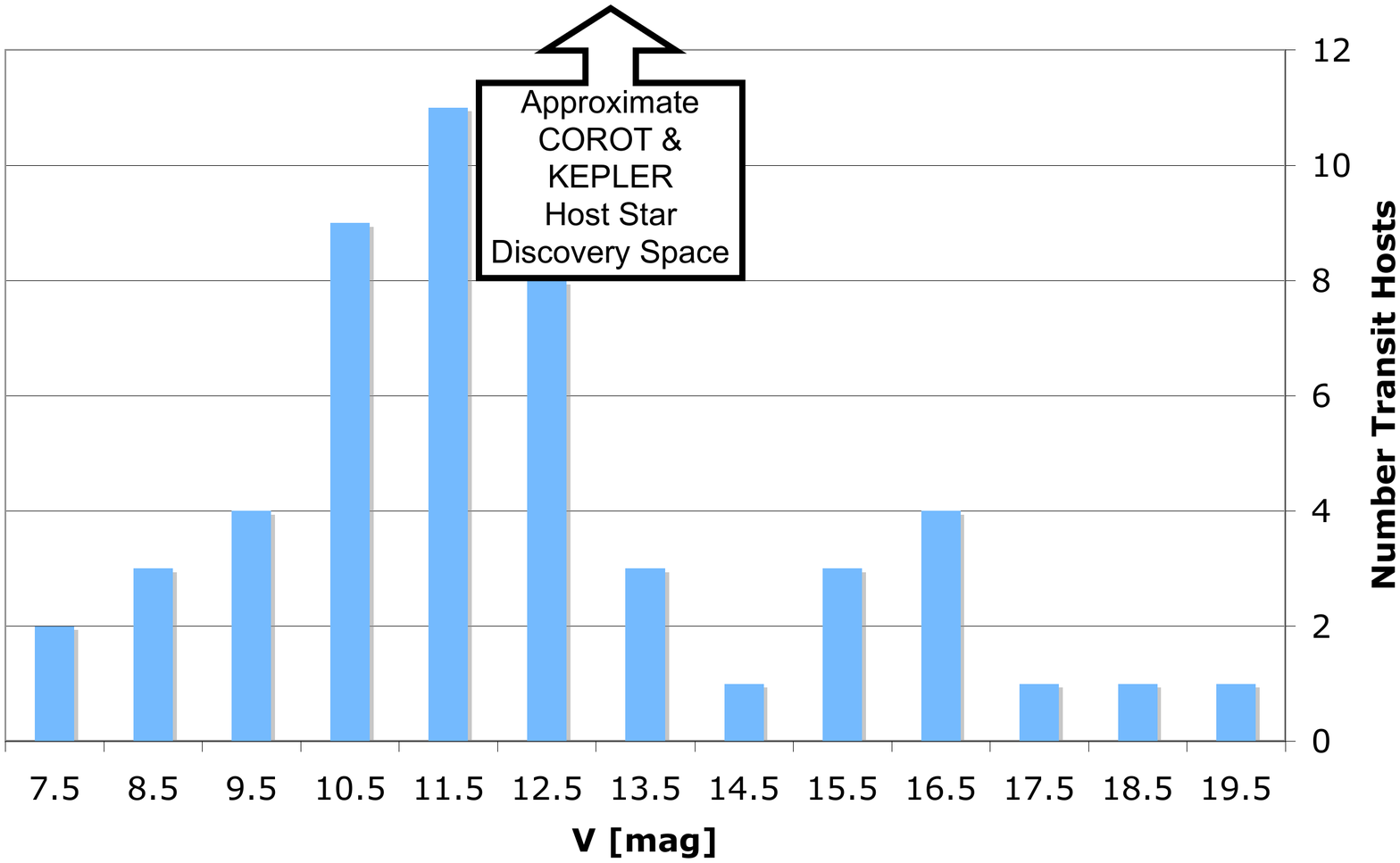} 
\vspace*{-1.75 cm}
 \caption{A histogram of the $V$ magnitude of host stars reported to have 
transiting planets.}
   \label{fig1}
\end{center}
\end{figure}

Most searches for transiting extrasolar transiting planets fall into two broad categories 
(see, eg, \cite{Mazeh08}): 
very wide-field searches targeting bright ($V < 12$) stars, and narrower field, 
pointed, dense-field observations monitoring fainter ($V > 14$) stars (Fig.~\ref{fig1}).  
Surveys of brighter stars have the advantage of more efficient spectroscopic follow-up 
due to the larger fluxes of their candidates.  Fainter searches are typically more 
efficient in the discovery phase, using less telescope time to densely monitor a 
similar number of dwarf star targets.

The space-based CoRoT (\cite{Bargeetal08a}) and 
KEPLER (\cite{Boruckietal08}) missions bridge part of this gap, pointing 
at particular dense stellar fields at low Galactic latitude, with the expectation 
that most of their prime target stars will have $12 < V < 14$.  This middle range is 
their "sweet spot" because the stellar mass function ensures that 
most dwarf hosts will lie at the faint end, 
while the best photometry required to search for small planets is achieved at the bright end.
Certainly the transiting planets reported by the CoRoT team to date, all with 
Jovian sizes, have host stars in this magnitude range (\cite{Bargeetal08b}; 
\cite{Alonsoetal08}; \cite{Aigrainetal08}).

Stars of this magnitude and fainter are increasingly more likely to be blended 
with foreground or background stars of similar brightness.  Furthermore, the ability to 
perform spectroscopic tests to rule out the possibility that a blended eclipsing 
binary is masquerading as a transiting planet becomes increasingly difficult as 
the host star becomes fainter.  
Since Jovian-sized planets generate a $\sim$1\% dip in host brightness when 
transiting a Sun-like star, a (totally) eclipsing stellar binary (EcB) system can be nearly 
five magnitudes fainter than a random blended neighbour along the line-of-sight 
and still generate a Jovian-like transit signal against the bright blended composite.  
For the same reason, surveys for 
terrestrial-sized planets must be able to exclude possible blended EcB 
up to the {\it ten\/} magnitudes fainter than their survey targets.  Clearly, 
this is an issue that the CoRoT and KEPLER teams will have to face 
in confirming planets whose apparent masses are too small to yield a  
robust radial velocity signature.

\section{The Lupus Survey and Lupus-TR-3b}

We were led to consider the issue of possible confusion with an EcB in follow-up 
work directed at the prime planetary transit candidate, Lupus-TR-3b
(\cite{Weldrakeetal08}), found in our initial deep survey of the Lupus 
region.  Like the CoRoT ({\it Serpens Cauda}: $l = 30, \, b \sim 0$; 
{\it Monoceros}: $l = 215, \, b \sim 0$) and KEPLER ({\it Cygnus}: $l = 76, b = 13$) 
fields, our Lupus survey was performed in a dense Galactic field ({\it Lupus}: 
$l = 331, b = 11$). 

In our survey, we monitored about 110,000 stars over 
a 0.66 square degree field in Lupus for 
53 nights in June of 2005 and 2006 with the ANU 40-Inch Telescope 
equipped with a wide field imager at Siding Spring Observatory (SSO) in 
Australia (\cite{Weldrakeetal07}; \cite{Baylissetal08b}).  The resulting 1783 exposures 
produced photometry to a precision of better than 0.025~mag (rms) 
for $\sim$16,000 stars.  The photometry was produced using 
an image subtraction technique, followed by SYSREM (\cite{Tamuzetal05}) to remove 
systematics (red noise) common to a large number of stars in 
the field.  The BLS detection scheme of Kov\'acs et al. (2002) was then used  
to identify promising candidates for the host stars of transiting 
planets.  This initial two-year survey is being extended in 
time to yield the SuperLupus Survey (\cite{Baylissetal08a}), 
will increase the 
sensitivity to longer period transiting planets.

The initial photometric selection process produced 
six candidates in the Lupus field, whose basic characteristics 
are listed in Table~\ref{tab1}.  More details will be 
forthcoming in a subsequent publication (\cite{Baylissetal08b}). 
All candidates were detected at a high level of 
significance, with a large number of in-transit 
photometric measurements, and depth-duration-period 
properties, as measured by the $\eta$ diagnostic 
(\cite{TingleySackett05}), that made them plausible transiting planet 
signatures.  

\begin{table}
\vskip 0.5cm
  \begin{center}
  \caption{Characteristics of Lupus Survey Candidates}
  \label{tab1}
  \begin{tabular}{|l|c|c|c|c|}\hline 
{\bf ~~Host Star} & {\bf ~V [mag]~} & {\bf ~Detection S/N$^ 1$~} & {\bf ~\# In-transit Points~} & {\bf ~$\eta$ Diagnostic$^ 2$~} \\
\hline
~~Lupus-TR-1~~    & 14.6	& 20.9	& 70	& 1.3 \\
~~Lupus-TR-2~~	& 14.8	& 26.3	& 59	& 1.0 \\
~{\bf Lupus-TR-3}~	& {\bf 17.4}	& {\bf 24.2} & {\bf 125} & {\bf 0.7} \\
~~Lupus-TR-4~~	& 16.3	& 23.2	& 141	& 1.2 \\
~~Lupus-TR-5~~	& 17.4	& 13.6	& 66	& 0.8 \\
~~Lupus-TR-6~~	& 17.4	& 15.0	& 57	& 0.6 \\
\hline
  \end{tabular}
 \end{center}
\vspace{1mm}
 \scriptsize{
 {\it Notes:}\\
  $^1$ The detection S/N is a measure of the significance of the size of collective transit signal.\\
  $^2$ The $\eta$ diagnostic is unity or below for likely planetary transits (see Tingley \& Sackett 2005).}  
\end{table}

After further scrutiny and 
follow-up observations, however, only Lupus-TR-3 
remained as a plausible candidate (Bayliss et al, 2008, 
in preparation).  The K1 dwarf exhibits a 1.3\% dip of about 
2.6~hour duration every $P = 3.91405$~days.  Subsequent radial velocity measurements 
taken with the MIKE echelle spectrograph on Magellan I 
displayed a confirming radial velocity signature 
of $K = 114 \pm 25$~m/s, appropriately in phase with the transit.
The planet, Lupus-TR-3b, thus has derived parameters of $M_p = 0.81 \pm 0.18 M_J$, 
$R_p = 0.89 \pm 0.07 R_J$, yielding a quite Jovian-like density of 
$\rho_p = 1.4 \pm 0.4$~gm/cm$^3$ (Weldrake et al.\ 2008).

This might have been the end of the story, had we not simultaneously been 
pursuing image deconvolution as a method to discover possible close neighbours 
in the vicinity of promising candidates.

\section{A (De)Convoluted Cautionary Tale}

The long train of monitoring images used to discover Lupus-TR-3b were 
obtained at SSO, a site with only moderate atmospheric seeing conditions.  
Furthermore, the ANU~40-Inch Telescope is known to have guider issues, and 
so the 5-minute frames were taken unguided.  As a result, a typical 
image would have a stellar point spread function (PSF) 
with a full width at half maximum of FWHM~$\approx 2"$.  Although this 
is considerably better image quality than obtained by wide-field 
transit surveys against bright stars, we decided to apply image 
deconvolution to the immediate Lupus-TR-3 field as an experiment aimed 
at the possibility of finding a line-of-sight "neighbour" within the 
typical FWHM.

For this purpose, we used a package originally 
developed for use in extragalactic lensing, but 
later optimised for point sources.  This algorithm, DECPHOT 
(\cite{Gillonetal06}; \cite{Magainetal07}), weights the PSF properly in the statistical sense, 
and uses all of the information on the image to 
achieve the best deconvolution.  On the other hand, it 
is CPU intensive, so that it is generally applied only to 
"postage stamp" areas around the immediate object of 
interest rather than to a complete CCD mosaic.

\begin{figure}[t]
\vspace*{0.5 cm}
\begin{center}
 \includegraphics[width=4.0in]{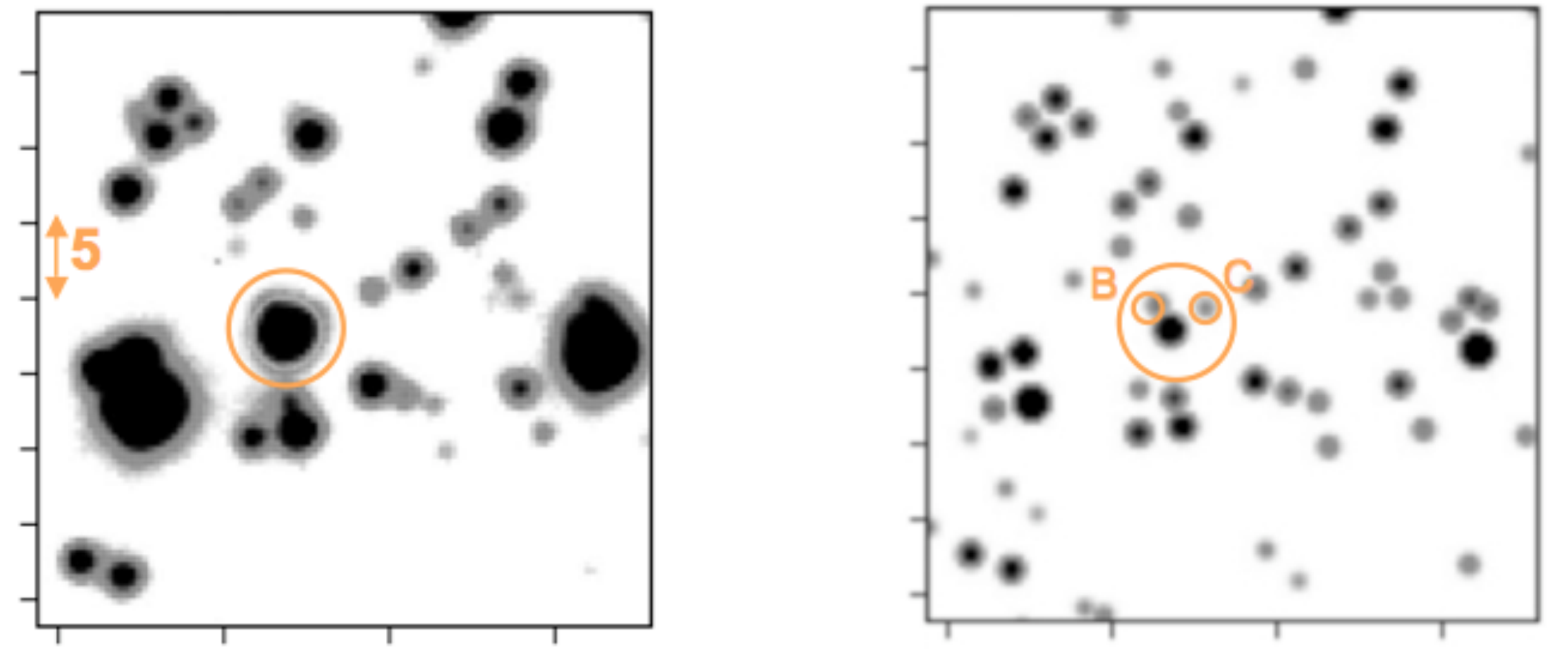} 
 \caption{Left: A typical image from the ANU~40-inch, for which 
 Lupus-TR-3 is the circled object.  Right: The same SSO field, to the 
 same scale, shown after application of DECPHOT.  Two neighbours, 
 stars $B$ and $C$, were indicated.}
   \label{fig2}
\end{center}
\end{figure}

An initial DECPHOT application to our SSO 
images revealed two possible neighbours, 
dubbed stars $B$ and $C$ (Fig.~\ref{fig2}).  Star $C$, 
located 2.2$''$ from Lupus-TR-3, 
is 5.6 $V$ magnitudes fainter than the target, and 
thus could not be responsible for the transit signature 
even if it was an EcB that fully eclipsed its primary.  
Star $B$, 1.7$''$ from the target is 4.1 $V$ magnitudes 
fainter, and could thus, in principle, be a planet-pretender.

However, by performing photometry on a time-sequence of DECPHOT 
deconvolved SSO images covering both in- and out-of-transit epochs, 
we were able to demonstrate that Lupus-TR-3 was undergoing a 
partial eclipse, not Star $B$ (\cite{Weldrakeetal08}).

Again, the story might have ended, had we not obtained a good seeing  
(FWHM $\sim 0.7''$) image of the Lupus-TR-3 field in the $Y$ band ($\sim$1.1 micron) 
with the {\sc PANIC} infrared imager on Magellan II, kindly provided 
by the Las Campanas Observatory staff.  Our aim was to check the 
positions and relative brightness of neighbours predicted by the 
deconvolution of SSO images.

\begin{figure}[b]
\vspace*{0.25 cm}
\begin{center}
 \includegraphics[width=3.0in]{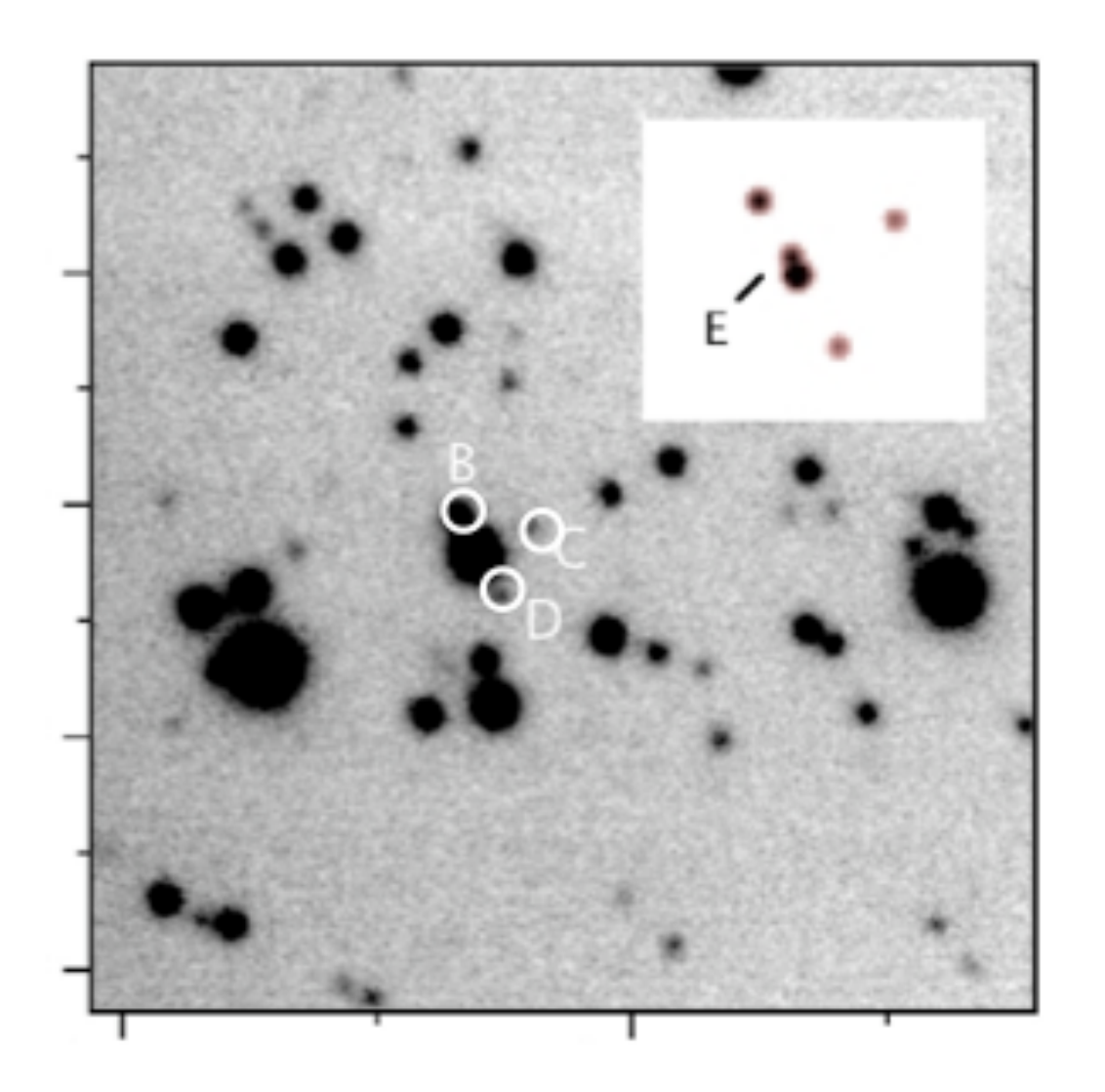} 
 \caption{Main: A Magellan $Y$-band PANIC image of the Lupus-TR-3 field 
 reveals stars $B$ and $C$, both predicted by DECPHOT analysis of 
 SSO images, and also a very faint neighbour $D$.  
 Insert:  Subsequent deconvolution of PANIC frames then predicted a 
 very near neighbour, labelled $E$, a possible, but unlikely, planet pretender.}
   \label{fig3}
\end{center}
\end{figure}

The Magellan $Y$ band image (Fig.~\ref{fig3}) indeed indicated 
that DECPHOT had correctly identified blended neighbours $B$ and $C$, 
but also indicated a third star, $D$, 1.8$''$ from the target that 
was 4.3 $Y$ magnitudes fainter than the Lupus-TR-3, and fainter still 
in $V$ --- too faint to be a culprit.  Naturally curious, however, 
we then applied DECPHOT to the high-quality PANIC image.  To our surprise, the 
deconvolution predicted yet another neighbour, this one, star $E$, 
2.8 magnitudes fainter in $Y$ than the target, and only 0.4$''$ distant. 
The inability of DECPHOT to identify it in our broad $V+R$ images from 
Siding Spring seemed to indicate that star $E$ must have $V > 21$, and 
thus would have to suffer an eclipse of 36\% or more to create a 
spurious transit signal.  Its faintness, coupled with the unusually 
parameters required for an EcB to mimic our signal (see Weldrake et al 2008), 
make star $E$ an unlikely, if possible, false positive.

Nevertheless, in order to check this possibility, and to provide yet 
another test of DECPHOT predictions, we applied for PANIC imaging 
time on Magellan over two epochs that would cover both in- and out-of-transit 
points.  The aim was to perform photometry on a time-series of deconvolved 
PANIC images capable of separating the original target star from 
the putative neighbour $E$ in order to directly discern, as we had earlier 
for star $B$, which was undergoing a partial eclipse.

\begin{figure}[t]
\vspace*{0.25 cm}
\begin{center}
 \includegraphics[width=4.0in]{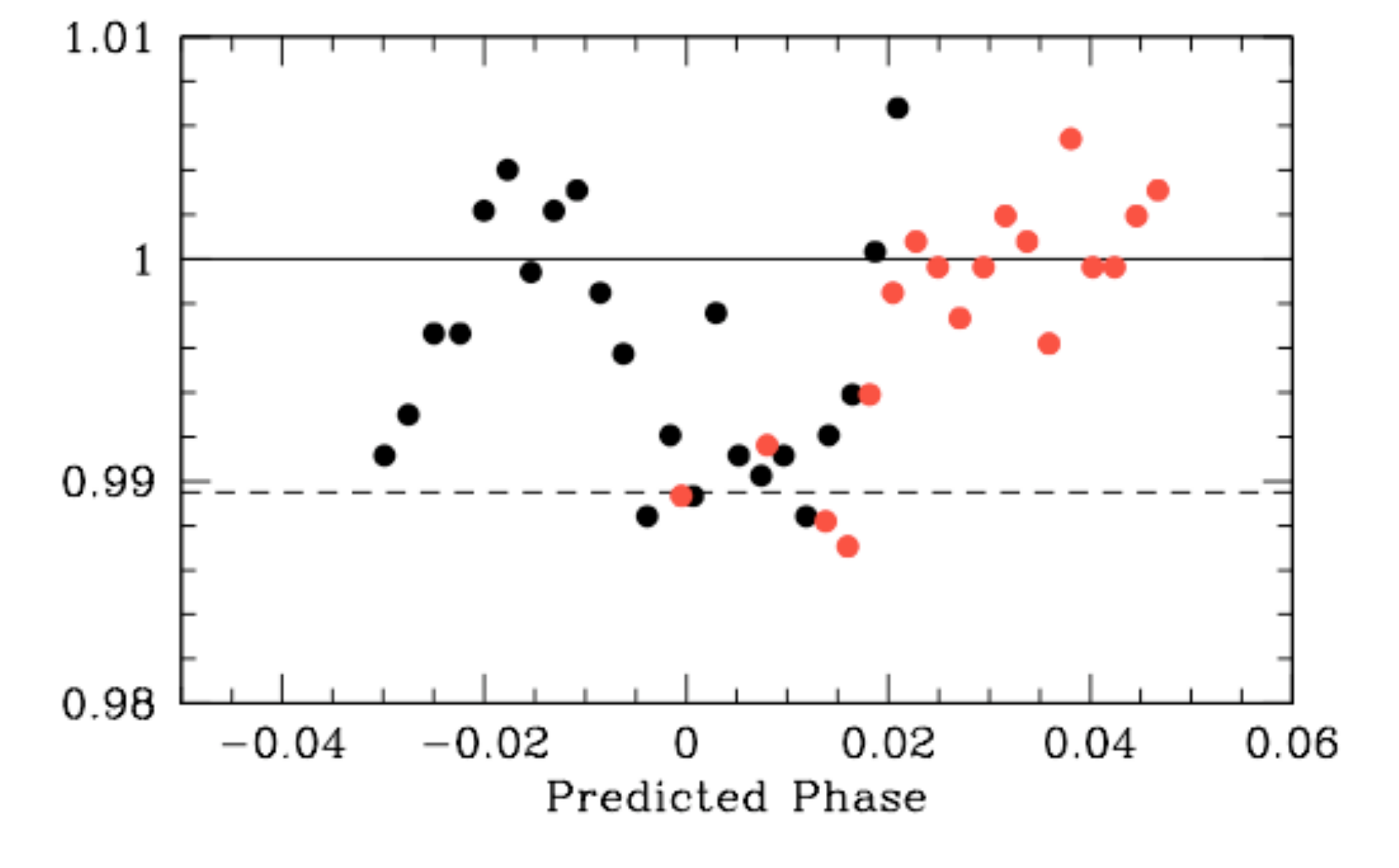} 
 \caption{Binned, ``on-mountain'' aperture photometry on PANIC near-IR images 
 of Lupus-TR-3.  The first two points were affected by "poor" Las Campanas 
 seeing of 1$''$.  Black (darker) points were taken in April 2008, and red 
 (lighter) points were obtained two periods later in May.}
   \label{fig4}
\end{center}
\end{figure}

These data were obtained in April and May of 2008 by one of us (PDS), 
producing over 200 frames in $Y$-band with a typical PSF of 0.55$''$.  
A very rough reduction on the mountain (see Fig.\ref{fig4}) indicates that 
we have captured the transit to within the uncertainties on the transit timing 
predicted a year earlier.

After a first data reduction, small (250 $\times$ 250 pixel) stamps around 
Lupus-TR-3 were processed (by MG) using DECPHOT, using only the best 
of the April 2008 data, namely those with seeing below 0.5$''$.  (Note that
the images are still well-sampled, as PANIC has pixels that translate into 
0.125$''$ on the sky.)  These data represent about 50\% improvement in 
image quality compared to our earlier PANIC snapshot of the field.

\begin{figure}[t]
\vspace*{0.25 cm}
\begin{center}
 \includegraphics[width=3.0in]{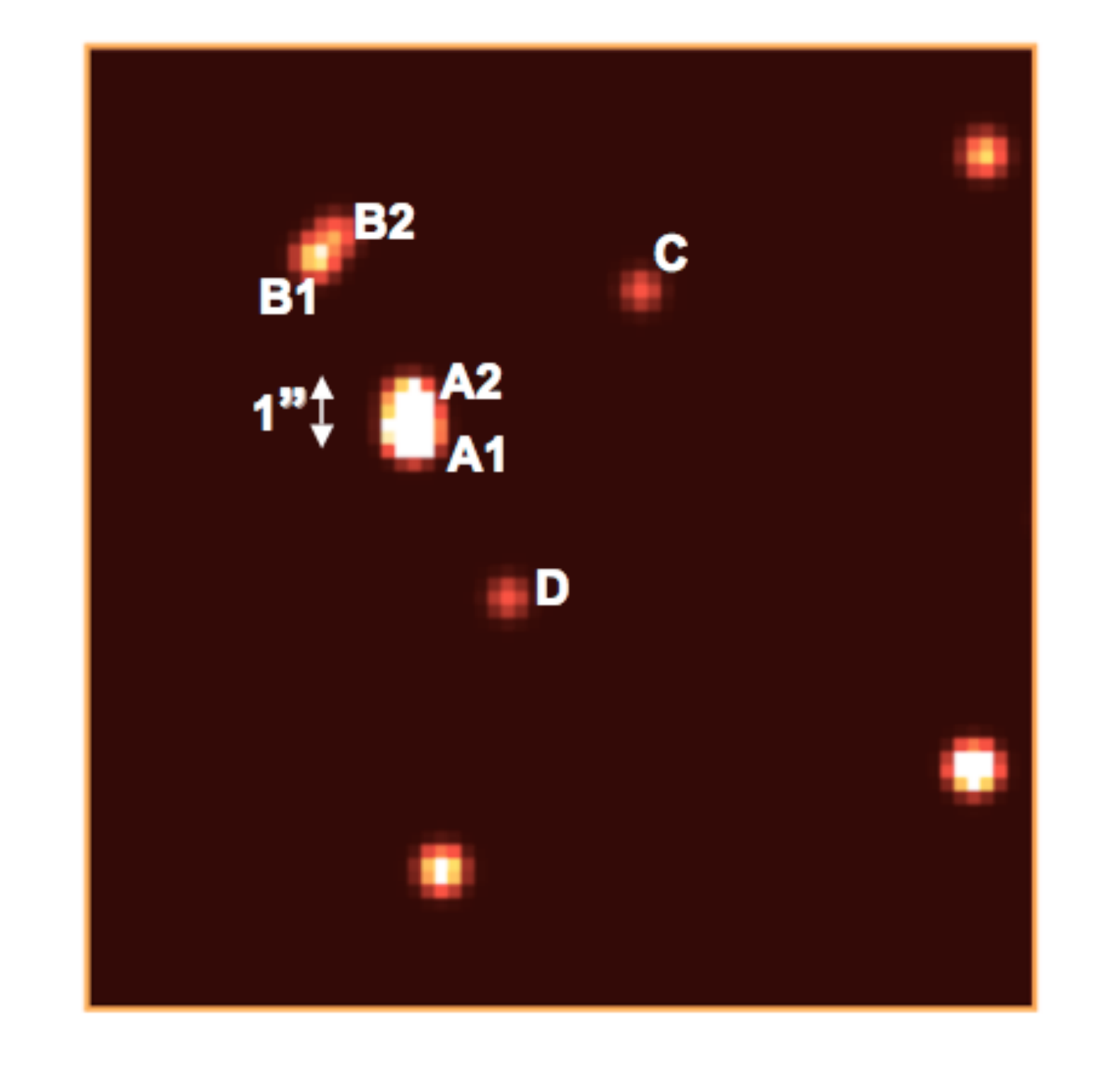} 
 \caption{Preliminary DECPHOT analysis of 
 a subset of the PANIC frames of the Lupus-TR-3 field reveal the 
 expected faint stars $C$ and $D$, but also that the (non-eclipsing) 
 star $B$ is actually composed of two components, as is the primary target.
 The length of the arrow denotes 1$''$ on the sky, indicating that this image is 
 considerably zoomed compared to those in Figs.~\ref{fig2} and \ref{fig3}.}
   \label{fig5}
\end{center}
\end{figure}

The results are surprising.  While neighbours $B$, $C$ and $D$ are 
all seen at their DECPHOT predicted positions and brightness, it 
appears that (non-ecliping) star B is actually two stars, here labelled $B1$ and $B2$ 
(see Fig.~\ref{fig5}).  While it is unclear whether the faint, close  
neighbour $E$ originally predicted from DECPHOT analysis of our 
earlier, somewhat poorer PANIC frame is actually present, it does seem  
clear that the primary target, Lupus-TR-3 itself has two components, 
$A1$ and $A2$ !  The component $A2$ is only 2.3 times fainter than $A1$ 
according to our preliminary analysis, and the two are separated by 
only 0.25$''$ on the sky, making it conceivable that they form a wide, 
physical binary.  Our original Lupus-TR-3 target, then, is composed of 
at least six, and perhaps seven, individual objects.

We are now in the process of undertaking a more detailed, thorough 
study of all the PANIC frames from Magellan of the Lupus-TR-3 field 
in order to understand 
whether (1) star $E$ is present, and (2) star $A1$, $A2$ or possibly 
$E$ is eclipsing.  Either $A1$ or $A2$ could host a transiting planet, 
albeit one of different planetary characteristics than were derived 
under the assumption that they formed a single star (\cite{Sackettetal08}).  
In the meantime, we offer our experience as 
a cautionary tale for all those who might imagine that the target 
they are observing, especially in fields of low Galactic 
latitude, is a single star 
with no faint, possibly EcB, blended neighbours, capable of mimicking 
a planetary transit.

\section{Conclusions}

Our experience with deep, high-spatial resolution imaging of the 
field of the planetary transit host star Lupus-TR-3 suggest that:
\begin{itemize}
\item It is vital to constrain the presence of blended line-of-sight 
	neighbours up to 5 magnitudes fainter than stars thought to 
	host Jovian-sized transiting planets, and up to 10 magnitudes 
	fainter than those thought to host terrestrial-sized transiting 
	planets. Faint, eclipsing stellar binaries may otherwise 
	cause an unacceptable level of false positives in searches for
	transiting extrasolar planets.	
\item Fields of low Galactic latitude ($b \sim 10$~deg or less) have a relatively 
	high probabilities of such impostors and confusing contaminants.
\item Even faint, undetected, {\it non\/}-EcB neighbours will influence 
	the inferred radius of detected planets, by causing an underestimation  
	of the true transit depth, and thus an underestimation of the 
	true radius.
\item The CoRoT and KEPLER missions may be particularly 
	susceptible to this contamination in their search for terrestrial-sized 
	planets.
\item Image deconvolution coupled with high-spatial resolution imaging is a powerful 
	tool for locating and studying many of these annoying neighbours.
\end{itemize}

\vglue 0.05cm

\end{document}